# Bandwidth-tunable thermal emitter via dynamic resonator coupling and loss

Jin-Woo Cho[1,*], Tanuj Kumar[1], Joseph A. Andrade[2], Shenwei Yin[1], Samir Rosas[3], Filiz Yesilköy[3], Dung T. Quach[2], Albert G. Wright[2], Colin M. Hessel[2], David Woolf[2], and Mikhail A. Kats[1]

[1]*Department of Electrical and Computer Engineering, University of Wisconsin-Madison, Madison, Wisconsin 53706, USA*
[2]*Physical Sciences Inc., Andover, Massachusetts 01810, USA*
[3]*Department of Biomedical Engineering, University of Wisconsin-Madison, Madison, Wisconsin 53706, USA*

**Corresponding author: cho252@wisc.edu*

## Abstract

We demonstrate a dynamically tunable omnidirectional thermal emitter that can be switched from a narrowband emission state to a broadband state. The structure is a metal-backed multilayer with subwavelength thickness across the mid-infrared, which forms an angle-insensitive (up to 60°) optical cavity with one input port. We used an ultrathin $VO_2$ layer to dynamically switch the coupling and loss coefficients of this optical cavity, resulting in narrowband emission from the vibrational mode of a polar dielectric layer that is part of the multilayer stack when the $VO_2$ is in its insulating phase, and broadband emission from a nearly impedance-matched cavity when the $VO_2$ is switched to its lossy metallic phase. We fabricated the structure around a free-standing silicon membrane and integrated it onto a printed circuit board (PCB), which enabled electrical switching of the emitter at the ten-to-hundred millisecond scale, and can be used for free-space identification, tagging, or communication.

## Introduction

Thermal emission is most-often spectrally broadband, incoherent, and spatially diffuse [1]. Meanwhile, many emerging infrared applications, including infrared camouflage [2], [3], [4], [5], radiative cooling [6], [7], [8], [9], and free-space communication [10], [11], [12] require precise control over the emission spectrum [2], [4], [5], [6], [8], [9], [10], [11], [12] and directionality [3], [7], which may also need to be dynamically tunable [4], [5], [8], [9], [10], [11], [12]. Dynamic control of the emission spectrum and directionality can be achieved via the integration of phase-change materials [4], [5], [8], [9], [10], [12], whose optical properties can be actively tuned by thermal [4], [8], [9], electrical [10], [12] or optical stimuli [5].

In this paper, we aim to design and demonstrate a tunable thermal emitter that switches between narrowband and broadband emissive states, but in both cases remains omnidirectional. Such switching behavior can be useful for infrared tagging, free-space communication, and tunable radiative thermal management.

We demonstrate that two emission states—one broadband and one narrowband—can be engineered in a single structure derived from the Salisbury screen [13], with the absorbing/emitting layer comprising a bilayer of a polar-oxide dielectric and an ultra-thin film of vanadium dioxide ($VO_2$), which has tunable infrared properties due to $VO_2$'s insulator-to-metal transition (IMT) [14]. **Fig. 1a** shows several different types of Salisbury-like geometries that incorporate $VO_2$, which can provide varying levels of amplitude and bandwidth tunability by coupling the tunable optical properties of $VO_2$ with a combination of optical cavity and material resonances.

Because the general optical features of $VO_2$ are broadband across the mid infrared (**Fig. 1b**), narrowband behavior must come from some combination of an optical cavity resonance and a narrowband resonance in some other constituent material. We specifically consider vibrational resonances of polar dielectrics such as $SiO_2$ and $Si_3N_4$ [15], which can then be coupled to an optical cavity, resulting in narrowband emission from the overall structure when the $VO_2$ is in its insulating phase. If the total stack thickness is significantly smaller than the operating wavelength (as is the case in all structures in **Fig. 1a**), the interference condition within the cavity varies little with the incidence angle [16], and the narrowband emission peak is expected to be nearly angle insensitive. In the metallic state, the ultrathin $VO_2$ film can contribute most of the loss and become the dominant emitting layer, resulting in broadband emission.

We realize a bandwidth-tunable thermal emitter using thin films of $VO_2$ and $SiO_2$ in a Salisbury-screen geometry (**Fig. 1a-iv**), resulting in switching from a narrowband state ($\lambda$ = 8 to 10 μm) to a broadband state ($\lambda$ = 6 to 14 μm) via Joule heating from an integrated electrical heater. The entire structure is suspended to reduce its thermal mass, achieving switching times of tens of milliseconds.

**Layer-by-layer engineering of amplitude and bandwidth tunability**

Starting from the simplest geometry in **Fig. 1a**, (i) a single $VO_2$ layer (a few hundred nanometers thick) directly on top of a metal reflector exhibits low absorptivity in the insulating phase and therefore low emissivity by Kirchhoff's law. With the $VO_2$ in the metallic phase, the increased optical loss (**Fig. 1b**) leads to higher emissivity. However, the metallic-phase $VO_2$ film here is optically thick (much thicker than its penetration depth, $\delta = \lambda/4\pi\kappa$) and has a large impedance mismatch with free space, producing substantial front-surface reflectance that also limits its emissivity. As a result, the emissivity contrast between the two phases is limited. The emission remains broadband in both phases because this structure has no resonance.

Introducing a transparent spacer between the $VO_2$ and the mirror creates a $VO_2$-Salisbury cavity that acts as a lossy asymmetric Fabry-Pérot (FP) resonator (**Fig. 1a-ii**). When the spacer thickness approximately satisfies the quarter-wave condition ($t_{\text{spacer}} = \lambda/4n_{spacer}$, where $n_{\text{spacer}}$ is the spacer index), the cavity is on resonance, enhancing absorption (and therefore the emissivity) in the lossy layer. With the $VO_2$ in the metallic phase, the cavity produces a peak in the emissivity. In its insulating phase, the low optical loss of $VO_2$ results in little absorption, so the emissivity remains low even at the cavity resonance.

However, the strong, spectrally flat loss of the metal-phase $VO_2$ (**Fig. 1b**) keeps the quality factor of the cavity low. Note that $VO_2$ is termed a bad metal because its real permittivity ($Re\{\epsilon_r\} = n^2 - \kappa^2$) remains positive over λ = 4–16 μm (**Fig. 1b**), unlike conventional metals such as gold, silver, or nickel, which have $Re\{\epsilon_r\} \ll 0$. The film thus behaves optically as a highly lossy dielectric more-so than a metal, similar to, e.g., a germanium thin film in the visible range [16]. However, in the insulating phase the cavity holds no absorber that could convert its resonance into an emission peak.

Incorporating a polar dielectric material (e.g., oxide or nitride) into the cavity in **Fig. 1a-iii** can enable narrowband emission when the $VO_2$ is in its insulating phase. Polar dielectrics support narrowband vibrational resonances [15] and thus can act as narrowband absorbers. The emission at the resonance wavelength can be enhanced by matching the FP mode to this material resonance via the spacer thickness. For the $VO_2$ in its metallic phase, the emission of the stack is primarily governed by its broadband absorption. As a result, structure (iii) increases the bandwidth contrast between the two states, while the amplitude contrast is mostly maintained.

Reducing the $VO_2$ to a sub-penetration-depth thickness (i.e., far below δ) within this vibrational-mode-assisted cavity, structure (iv), maximizes both amplitude and bandwidth tunability (**Fig. 1a-iv**). The very small thickness of the $VO_2$ layer leads to a suppression of the front-surface reflection, allowing the incident field to penetrate into the underlying cavity. This transmitted light undergoes multi-pass absorption within the cavity. For the reverse process, this is equivalent to strong thermal emission (see also **Supplementary Section I**, and **Figs. S1** and **S2**). This enhanced emission remains broadband, because the strong and spectrally flat loss of the bad-metal $VO_2$ keeps the cavity quality factor low, as in structures (ii) and (iii). However, the suppressed front-surface reflection makes the broadband emission stronger than in structure (iii). In the insulating phase, the transparent $VO_2$ leaves the polar-material narrowband peak intact. The $VO_2$ phase transition therefore redistributes the dominant optical absorption in this structure from the polar layer to the $VO_2$ film, switching the emitter from narrowband to broadband.

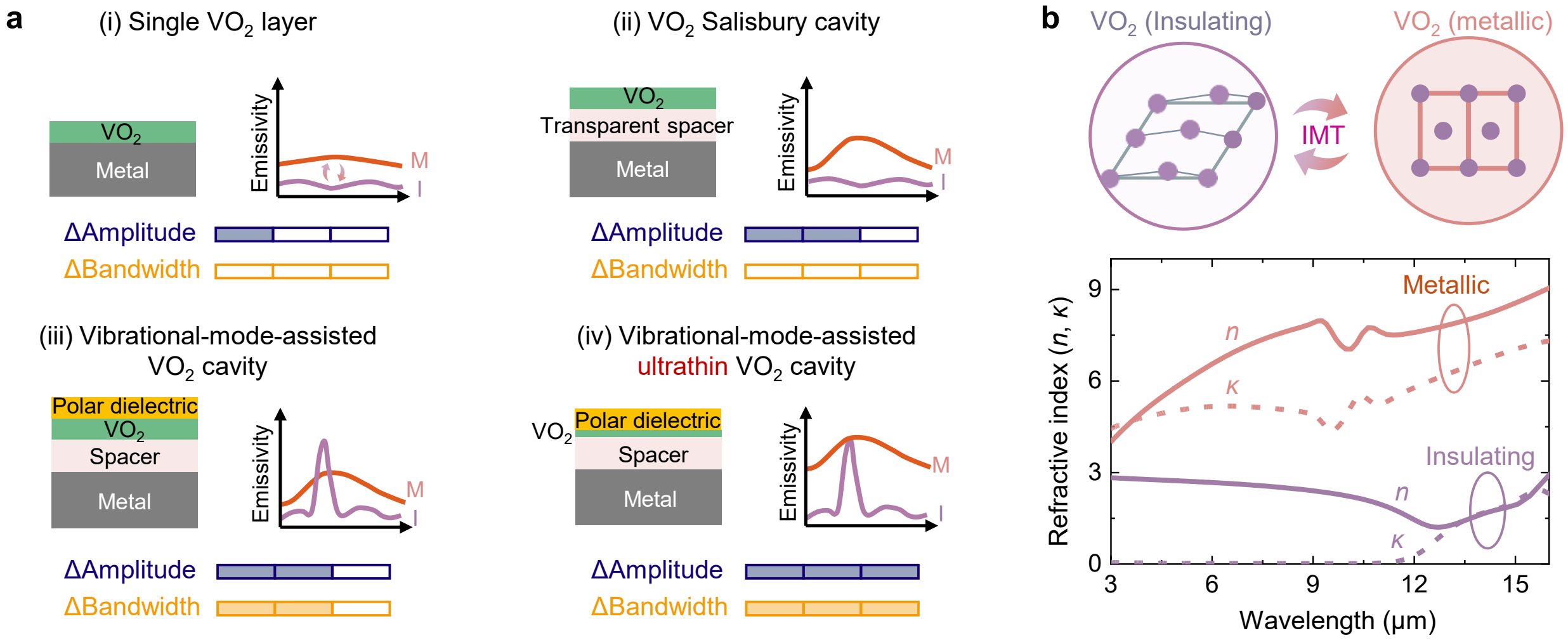

**Figure 1. $VO_2$-based dynamically tunable thermal emitter.** (**a**) Thin-film stack configuration and schematic emissivity spectra in the insulating (I) and metallic (M) phases for (i) a single $VO_2$ layer on a metal film, (ii) a $VO_2$-based Salisbury cavity, (iii) a vibrational-mode-assisted $VO_2$ cavity, and (iv) a vibrational-mode-assisted ultrathin-$VO_2$ cavity. Here the spacer is assumed to be an infrared-transparent material, and the polar-dielectric layer exhibits a vibrational resonance at its corresponding frequency. For each structure, the amplitude and bandwidth tunability (ΔAmplitude and ΔBandwidth) are qualitatively rated. (**b**) Schematic of atomic unit cells of $VO_2$ in its insulating and metallic phases with corresponding complex refractive indices of $VO_2$ in both phases [14].

## Design rules for narrowband and broadband emission in a single structure

The narrowband emission corresponding to the $VO_2$ insulating phase and the broadband emission corresponding to the $VO_2$ metallic phase are governed mostly by the spacer thickness ($t_{spacer}$), comprising silicon (Si), and the $VO_2$ thickness ($t_V$), respectively (**Fig. 2**).

When the $VO_2$ is in the insulating phase, the narrowband emission arises from the spectral overlap between the FP mode of the cavity and the vibrational resonance of $SiO_2$ near $\lambda \sim 9$ μm (**Fig. 2a**). The calculated emissivity (and absorptivity, via Kirchhoff's law) as a function of $t_{spacer}$ in **Fig. 2b** shows that the FP mode intersects the $SiO_2$ phonon response at $t_{spacer} \sim 620$ nm, producing a high-amplitude, spectrally selective thermal-emission peak at $\lambda \sim 9.4$ μm. The $SiO_2$ thickness is separately optimized to ~140 nm to maximize the peak emissivity. Because insulating $VO_2$ is low-loss and optically thin in this spectral range (**Fig. 1b**), it only weakly perturbs the cavity interference condition. The value of $t_V$ is set by the metallic-phase design (**Figs. 2d** and **2e**).

In the metallic phase, broadband emission arises from cavity-assisted absorption enabled by transmission through the ultrathin $VO_2$ layer (**Fig. 2d**). When $t_V$ is $t_V \ll \delta$, the incident field penetrates the $VO_2$ layer and couples into the cavity rather than being reflected at its front surface as would be the case for a thicker metallic layer. The cavity then enables several passes of the wave through the lossy $VO_2$ layer, enhancing its emission. Because the cavity loss is high (and hence the Q factor is very low), the emissivity spectrum is broadband (**Supplementary Section I**).

To better understand this regime, we can view the $VO_2$ layer as a shunt sheet element in a transmission-line model [17], with sheet impedance $Z_{VO2} = 1/(\tilde{\sigma} t_V)$, where $\tilde{\sigma} = -i\omega\epsilon_0(\tilde{\epsilon} - 1)$ is the complex optical conductivity of metallic $VO_2$ (**Supplementary Section II**). Because $Z_{VO2}$ scales inversely with $t_V$, the film thickness tunes the input impedance $Z_{in}$ of the full stack when the $SiO_2$ and spacer thicknesses are held constant. We calculate $Z_{in}$ as a function of $t_V$ from 0 to 100 nm over the mid infrared and, at each wavelength, identify the thickness that minimizes the impedance mismatch $|Z_{in} - Z_0|$, where $Z_0$ is the free-space impedance (377 Ω) (**Fig. 2e**). We find that over the 8–14 μm wavelength range of interest, the optimum thickness $t_V$ is between 0 and 40 nm, with $t_V \sim 20$ nm as a good compromise. Indeed, the average emissivity over λ = 8–14 μm, weighted by the 300-K and 340-K blackbody spectra, shows its maximum near $t_V \sim 20$ nm (**Fig. S5**). As a result, we choose $t_V \sim 20$ nm as the design to fabricate (**Fig. 3**).

The layer-resolved emissivity calculations in **Figs. 2c** and **2f** confirm that the dominant emission is from the $SiO_2$ layer when the $VO_2$ is in its insulating phase (**Fig. 2c**) but shifts to the metallic-phase $VO_2$ (**Fig. 2f**) after the phase transition. Despite the different mechanisms and bandwidths, both states of the structure maintain nearly angle-insensitive emissivity up to 60°. Notably, the high-index Si spacer keeps the internal propagation angle small for a broad range of incidence angles from air, and thus the cavity interference varies only weakly with the incident angle (**Figs. 2c** and **2f**).

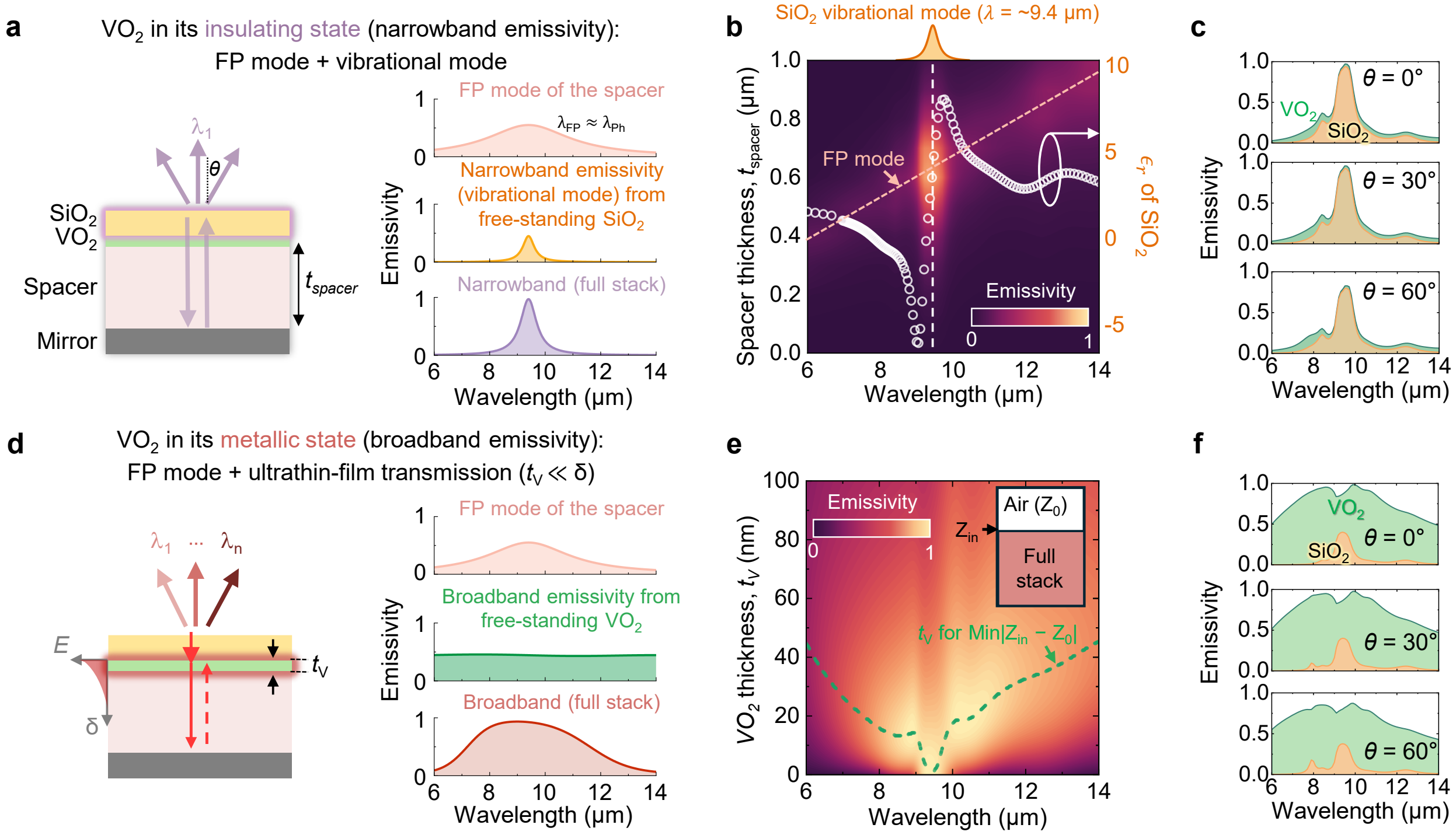

**Figure 2. Mechanism of the $VO_2$-based bandwidth-tunable thermal emitter.** (**a**) The vibrational-mode-assisted $VO_2$ cavity in the insulating phase, where $SiO_2$ serves as the vibrational resonance, resulting in narrowband thermal emissivity by combination of the Fabry-Pérot (FP) mode of the spacer and the vibrational mode of $SiO_2$. (**b**) Calculated emissivity spectra as a function of spacer thickness ($t_{spacer}$). (Right axis) The real part of the $SiO_2$ permittivity. (**c**) Calculated emissivity spectra at emission angles ($\theta$) of 0°, 30°, and 60° in the insulating phase, with the layer-resolved emissivity contributions. (**d**) The same structure as in (a) but in the metallic phase, resulting in broadband thermal

emission by combination of the FP-mode and transmission through the ultrathin lossy layer, where the thickness of $VO_2$ ($t_V$) is far below the penetration depth ($\delta$). (**e**) Calculated emissivity spectra as a function of $t_V$. The thickness minimizing the difference between the input impedance ($Z_{in}$) of the full stack and the free-space impedance ($Z_0$), Min$|Z_{in} - Z_0|$, is plotted on top of the emissivity map (dashed line). (Inset) Schematic showing $Z_{in}$ at the interface between air ($Z_0$) and the full stack. (**f**) Calculated emissivity spectra at emission angles ($\theta$) of 0°, 30°, and 60° in the metallic phase, with the layer-resolved absorption contributions. In (c) and (f), the oblique emissivity values are averaged between the s- and p-polarizations.

### Fabrication of the $VO_2$-based bandwidth-tunable thermal emitter

We fabricated the designed structure in **Fig. 2** by sequentially coating the layers onto a 500-nm-thick crystalline-Si (c-Si) membrane (NSUF1043E, Norcada), which serves as the spacer layer (**Fig. 3a**). The membrane platform was chosen for three reasons. First, high-quality $VO_2$ generally requires high-temperature processing (typically 400–650 °C for crystallization and stoichiometry control, regardless of the deposition method [14], [18], [19]), which is not readily compatible with the more-obvious bottom-up sequence of metal, spacer, and $VO_2$, for most material combinations. Instead $VO_2$ was synthesized via the sol-gel method on the c-Si surface [18], and the backside metal layers were deposited afterward at low temperature. Second, the small thermal mass of the membrane can reduce the power required for Joule heating and enable fast thermal switching. Third, the c-Si membrane itself is a low-loss, high-index ($n \sim 3.5$) spacer, enabling nearly angle-insensitive cavity interference conditions (**Figs. 2c** and **2f**).

A ~20-nm-thick $VO_2$ film was synthesized by Physical Sciences Inc. on the c-Si membrane by the sol-gel method [18], starting from a spin-coated $V_2O_5$ precursor followed by annealing at > 500 °C under a reducing atmosphere, which reduces $V^{5+}$ toward $V^{4+}$ to achieve the $VO_2$ stoichiometry. Then, e-beam evaporation was used to deposit $SiO_2$ on the $VO_2$ film and Ge, a thin (< 5 nm) titanium (Ti) adhesion layer, and gold (Au) sequentially on the backside of the c-Si membrane. The target spacer thickness in the design is ~620 nm, whereas the c-Si membrane we had available is ~500 nm thick. The Ge layer on the backside compensates for this deficit in optical path. Calculated emissivity as a function of Ge thickness (for fixed c-Si membrane thickness) shows that the maximum emissivity nearly saturates above 40 nm (**Supplementary Section III** and **Fig. S6**); we chose a Ge thickness of ~50 nm. The 80-nm-thick Au layer acts as the back-reflector. A photo of the fabricated membrane chip is shown in **Fig. 3b** (left). Microscope images show no visible wrinkles or cracks in the membrane area after the coating processes on both sides (**Fig. 3b**, right).

Micro-Raman spectroscopy was employed to identify the phase of the annealed $VO_2$ layer at 25 °C (insulating) and 90 °C (metallic), as shown in **Fig. 3c** (see **Methods**). Raman spectra (**Fig. 3c**) were measured through the $SiO_2$ top layer (**Fig. 3a**, right), and were calibrated using the optical-phonon mode at 520 $cm^{-1}$ of c-Si. The obtained spectra were averaged by scanning a 50 × 50 $\mu m^2$ area with a step size of 10 µm, with each point measured 10 times. The Raman spectra at 25 °C show modes at 198, 224, 271, 301, 338, 432, 628, and 825 $cm^{-1}$, characteristic of the monoclinic insulating phase of $VO_2$ (**Fig. 3c**, upper) [14]. The V–O stretching mode at ~628 $cm^{-1}$ is slightly blue-shifted from the bulk M1 phase (typically around ~615 $cm^{-1}$) [20], which is consistent with the compressive strain of thin $VO_2$ films [21]. Upon heating to 90 °C, $VO_2$ undergoes a phase transition from the insulating phase to the metallic phase, leading to the disappearance of all $VO_2$ Raman peaks (**Fig. 3c**, lower).

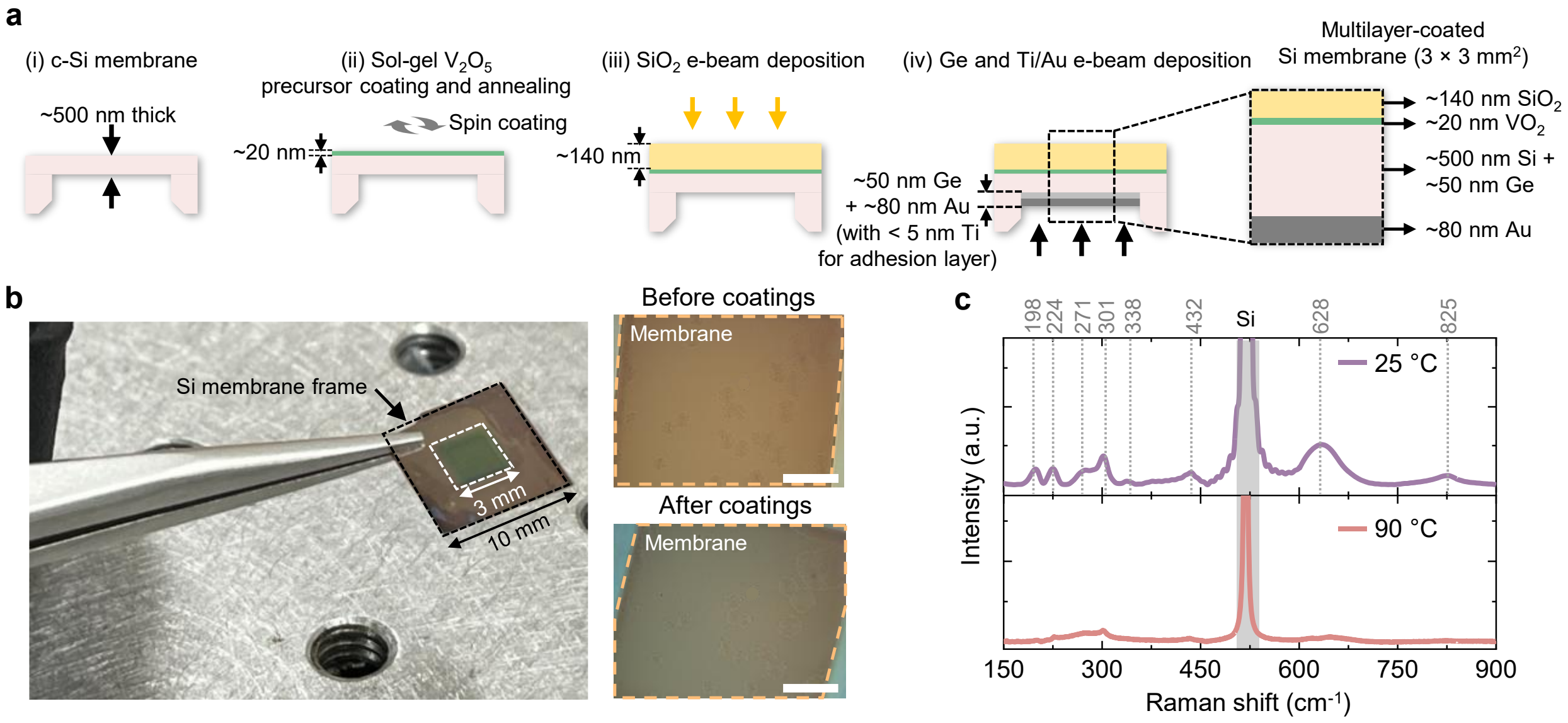

**Figure 3. Fabrication and material characterization**. (**a**) Fabrication process for the $VO_2$-based tunable thermal emitter on a c-Si membrane. (**b**) (Left) Photo of the fabricated structure. The white dashed square indicates the c-Si membrane area where the fabrication was performed. (Right) Microscope images (top) before and (bottom) after the complete coating process in (a), showing that the membrane remains flat and undamaged. Scale bars: 500 μm. (**c**) Raman spectra of the $VO_2$-based thermal emitters at (upper) 25 °C and (lower) 90 °C. Gray dashed lines indicate the peak positions of the insulating phase of $VO_2$, and the gray-colored area corresponds to the optical-phonon mode of the c-Si membrane (~520 $cm^{-1}$), used for spectral calibration.

## Electrically driven switching of the $VO_2$-based thermal emitter

To enable dynamic tuning, a gold microheater was deposited onto the membrane chip by e-beam evaporation through a shadow mask (**Fig. S7**). The chip was then transferred onto a printed circuit board (PCB), and the heater electrodes were connected to the PCB via aluminum wire bonding.

To characterize the emission switching, we measured emissivity via Kirchhoff's law (i.e., emissivity is equal to 1 – reflectivity) using an infrared microscope integrated with a Fourier-transform infrared (FTIR) spectrometer (**Fig. 4a** and see **Methods**) under electrical biasing. The current was swept from 0 to 32 mA in both increasing and decreasing directions. We show the emissivity spectra at 0 mA and 26 mA in **Fig. 4b**, corresponding to the insulating and metallic phases of $VO_2$. The spectra are nearly identical to those measured when controlling the temperature using an external heating stage (THMS600, Linkam); see symbols in **Fig. 4b** and **Fig. S8**. Note that these reference measurements using the external heater were performed prior to integrating the membrane chip onto the PCB.

In **Fig. 4c** we plotted the current-dependent emissivity for two wavelengths: $\lambda = 8.0$ μm, corresponding to the peak emissivity for $VO_2$ in the metallic phase, and $\lambda = 9.4$ μm, corresponding to peak emissivity when the $VO_2$ is in its insulating phase. The latter peak is due to the vibrational mode of $SiO_2$. A pronounced hysteresis is observed in the emissivity at $\lambda = 8.0$ μm during forward and reverse current sweeps, resulting from the intrinsic hysteresis in the $VO_2$ phase transition. In contrast, the emissivity at $\lambda = 9.4$ μm, where a hybrid optical mode emerges (**Fig. 2b**), is mostly unchanged as a function of current (**Fig. 4c**).

Together with the current-dependent emissivity, we recorded the voltage across the heater (right axis, **Fig. 4c**), which increases linearly with current, indicating that the Au heater resistance remains approximately constant throughout our experiments. Because the driving current primarily flows through the metal heater rather than the $VO_2$ layer, the biasing avoids the hysteretic resistive switching that can occur when forcing current through the $VO_2$ [22]. This configuration therefore enables stable and reproducible electrically driven switching of the emission states.

We characterized the angle-resolved emissivity for both the insulating and metallic phases of our emitter by performing angle-dependent reflectance measurements using a Woollam infrared variable-angle spectroscopic ellipsometer (IR-VASE) with a gold reference mirror (**Fig. 4d** and see **Methods**). The subwavelength multilayer design ensures angle-insensitive emission in both states, up to 60°. Because the IR-VASE instrument cannot measure near normal incidence, we obtained the normal-incidence emissivity of both phases from FTIR measurements (**Fig. 4b**). The experimental data matches the angle-dependent calculations using the transfer-matrix method (**Fig. 4e**).

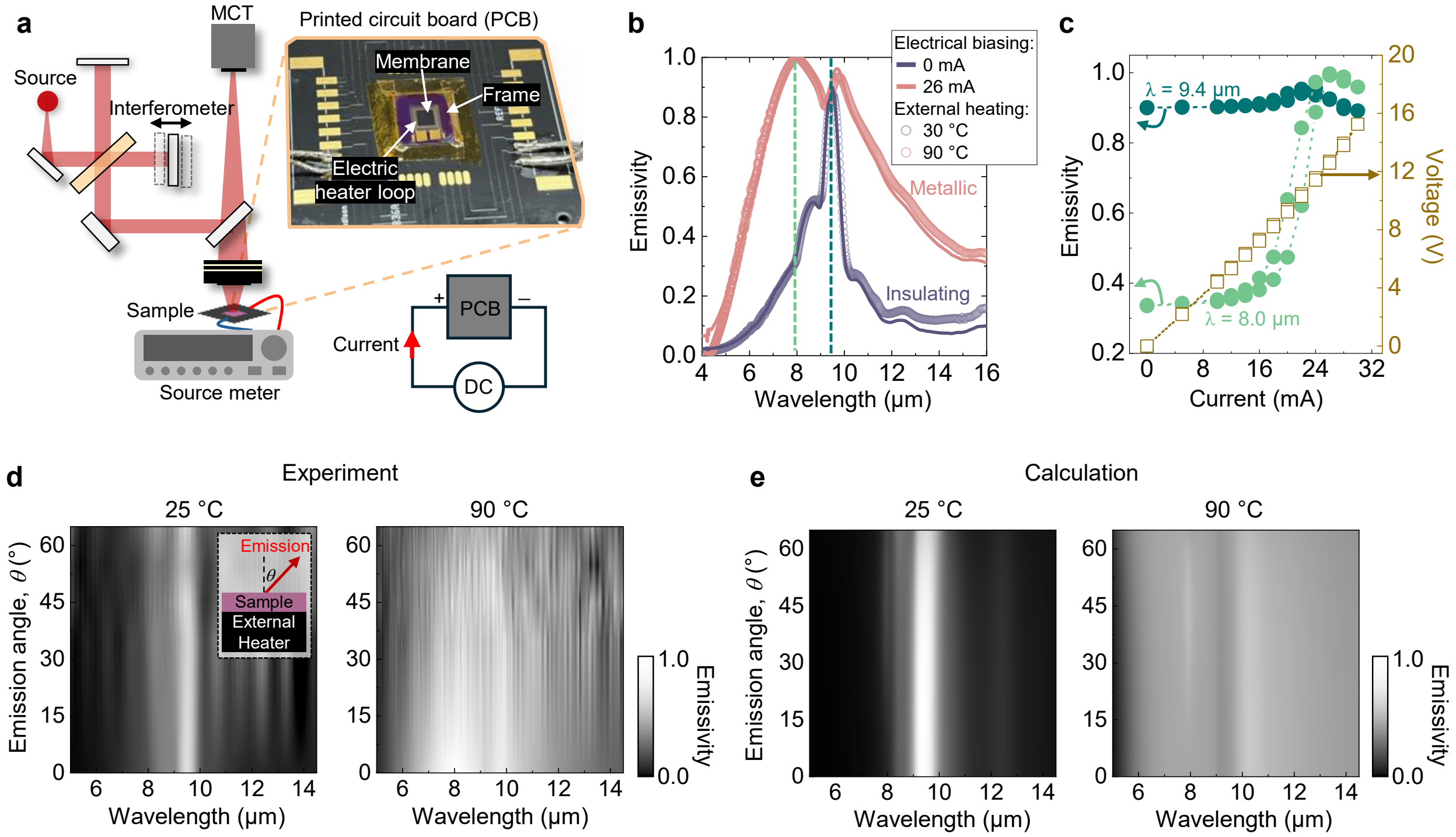

**Figure 4. Optical characterization of the fabricated electrically tunable thermal emitter**. (**a**) Schematic of our measurement setup, along with a photo of the membrane chip mounted onto a printed circuit board (PCB). (**b**) Measured emissivity spectra of the device, corresponding to $VO_2$ in its insulating and metallic phases, obtained via (lines) electrical biasing and (symbols) using an external heater. (**c**) (Left axis) Emissivity at λ = 8.0 μm and 9.4 μm, plotted together with (right axis) the applied voltage, as a function of current. (**d**) Measured and (**e**) calculated angle-resolved emissivity spectra for the two states.

## Infrared free-space communication demonstration

To characterize the temporal response of the device, we applied a current (26 mA) to switch the emitter, and monitored the apparent temperature using an infrared thermal camera (**Fig. 5a**). The apparent temperature changes with a turn-on time of at most ~90 ms and a turn-off time of at most ~120 ms; we believe that a 1/e switching time is slightly shorter, on the scale of tens of ms, but we cannot resolve the specific number due to the camera frame rate (i.e., 60 Hz). The turn-on time is governed by thermal diffusion from the heater to the membrane center, while the turn-off time is determined by the heat conduction from the membrane to the surrounding frame. For systems requiring faster temporal response, both could be shortened by optimizing the heat-path design, for example, by placing the heater closer to the emitter area or reducing the membrane dimensions for faster turn-on, and by introducing high-thermal-conductivity paths to the frame for faster cool down.

Using our tunable emitter, we demonstrated a simple communication protocol in Morse code, using dash (–) and dot (·) signals in the apparent temperature, as recorded by an infrared camera. In **Fig. 5b** and **Supplementary Movie S1**, we transmitted an "SOS" sequence. As the device was iteratively switched, the baseline temperature did not rise significantly, resulting in a nearly constant apparent temperature contrast throughout the transmission.

A thermal camera integrates the emission over its full operating spectral range and thus shows the device only as bright or dark. In contrast, a multispectral or hyperspectral camera resolves individual wavelengths and can distinguish differences in emissivity across the spectrum. We performed hyperspectral reflectance imaging of the same device used in **Figs. 5a** and **5b**. The result in **Fig. S9** (**Supplementary Section V**) shows that the emissivity at $SiO_2$'s vibrational mode (9.4 μm) remains high in both insulating (current at 0 mA) and metallic (26 mA) phases. However, the emissivity at 8.0 μm shows distinct changes across the IMT, consistent with the emissivity-*vs*-current in **Fig. 4c**.

Our bandwidth-tunable emitter can be extended beyond the single-channel (i.e., single resonance) demonstration to a dual-channel infrared label by patterning polar dielectrics with different vibrational resonances (**Fig. 5c** and **Fig. S10** in **Supplementary Section V**). As a conceptual example, the letters "UW" can be spelled out using a patterned $SiO_2$ layer (vibrational resonance at ~9.4 μm) and the letters "MADISON" can be spelled using a $Si_3N_4$ layer (vibrational resonance at ~12 μm) in **Fig. 5c**, while the background is the bare $VO_2$ film without any polar dielectrics. The patterned regions would show narrowband emission when the $VO_2$ is in its insulating phase, whereas the background would remain spectrally flat with low emissivity. Once the $VO_2$ transitions to its metallic phase, all regions would switch to broadband emission.

We expect that a thermal camera could hardly resolve such a pattern, whereas a multispectral or hyperspectral camera would see each region in its own resonance wavelength (in **Fig. 5c**, "UW" at ~9.4 μm and "MADISON" at ~12 μm). Heating the device above the transition temperature would then erase both patterns, because all regions transition to the broadband emission state. The label would thus be mostly invisible to a conventional LWIR camera, but readable to a multispectral or hyperspectral camera, and could be erased on demand, which may be useful for identification.

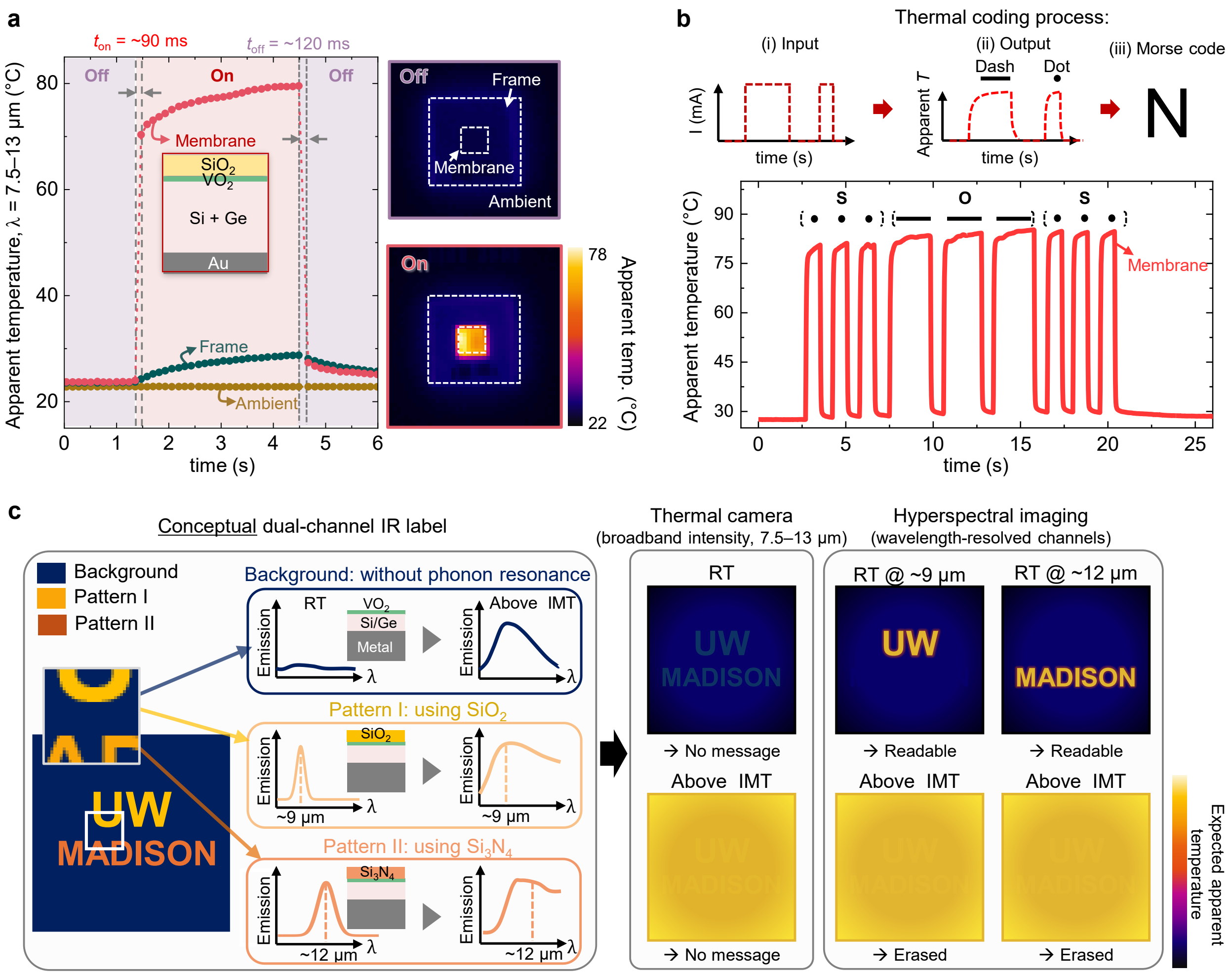

**Figure 5. Infrared free-space communication and identification.** (**a**) The apparent temperature of our tunable emitter (patterned around the Si membrane) measured using an infrared camera, before and after an electrical signal is turned on and off. The apparent temperature of the frame and ambient environment is also shown. (**b**) Infrared Morse-code demonstration. (Upper) Thermal coding process from the electrical bias input to the recorded apparent temperature. (Lower) Transmission of the "SOS" sequence recorded using an infrared camera. The apparent temperature was obtained from the membrane area, as in (a). (**c**) Conceptual schematic of a dual-channel IR label using two different polar dielectrics, such as $SiO_2$ and $Si_3N_4$. Using different vibrational resonances, spatial patterns at different wavelengths can be encoded when the $VO_2$ film is in its insulating phase, readable by a multispectral or hyperspectral camera but mostly invisible to a thermal camera. Heating above the $VO_2$ transition erases both patterns.

## Discussion

Our goal in this work was to create a thermal emitter that can be switched from a narrowband to broadband state—a functionality that can be used for free-space identification, tagging, or communication. To achieve tunability, we designed a metal-backed thin-film optical structure similar to a Salisbury screen, including an ultrathin layer of vanadium dioxide ($VO_2$)—a phase-transition material with an infrared absorption coefficient that changes significantly with temperature—and a polar dielectric layer to select an emission wavelength for the narrowband emission state. Our design ensures that we have excellent impedance matching between the structure and free space for both the narrowband and broadband optical states.

Because the entire thin-film stack is subwavelength and the operating wavelength of the narrowband state is partially selected by the polar dielectric, the emissivity in both states varies little with angle, up to 60° from the normal. The ability to dynamically switch between these two omnidirectional states makes the emitter a good label or tag that can be read at a distance using a multispectral or hyperspectral infrared camera.

We integrated the emitter on a printed circuit board (PCB), such that the temperature of the suspended tunable emitter could be easily changed with applied current. Using this integrated emitter, we demonstrated a free-space infrared communication scheme that can also be extended to multiple spectrally encoded channels for labeling and identification.

## Methods

### Raman spectroscopy measurements

Raman spectra were obtained using a λ = 532 nm laser as an excitation source, integrated with a Horiba LabRAM HR Evolution micro-Raman spectrometer equipped with a long working distance 50× objective (**Fig. 3c**). The Raman signal was calibrated using the phonon mode of a standard Si wafer at 520 $cm^{-1}$. To minimize heating of the $VO_2$ films by the source laser, the laser power was reduced to < 0.6 mW. The spectra were averaged over a 50 × 50 $\mu m^2$ area with a step size of 10 µm, with 10 acquisitions per point.

### Temperature-dependent measurements of reflectance and emissivity

The temperature-dependent emissivity spectra (i.e., not the measurements driven by current) were measured by applying Kirchhoff's law (i.e., 1− reflectance), from measurements using a Fourier-transform infrared (FTIR, Bruker Vertex 70) spectrometer connected to an infrared microscope (Hyperion 2000; objective NA = 0.4), using an Au mirror as reference, as shown in **Fig. 4a**. The temperature was controlled using a heating and cooling stage (THMS600, Linkam). The spectra were measured during both heating and cooling cycles in 5 °C intervals.

### Angle-resolved temperature-dependent measurements

Angle-dependent reflectance measurements using a Woollam infrared variable-angle spectroscopic ellipsometer (IR-VASE) were used to obtain the angle-resolved temperature-dependent emissivity spectra of the $VO_2$ full stack, when the $VO_2$ film is in its insulating (room temperature) and metallic (90 °C) phases (**Fig. 4d**). The spectra were measured for both s- and p-polarizations and averaged.

### Thermal imaging

Thermal images were captured using a FLIR A325sc camera (spectral range λ = 7.5–13 µm). The sample was heated by Joule heating through the micro-heater loop. The heating current was controlled from 0 to 26 mA using a source meter (Keithley 2612B). We set the camera emissivity to 0.95, which is the emissivity the camera assumes when converting measured thermal-emission power into an apparent temperature.

Hyperspectral imaging in **Fig. S9** was performed in reflection geometry using a quantum cascade laser-based discrete frequency IR microscope (DRS Daylight Solutions) integrating four tunable sources to cover the 950–1800 $cm^{-1}$ range with 2 $cm^{-1}$ spectral resolution. The device was illuminated with collimated, linearly polarized light at normal incidence, and its bandwidth was tuned via electrical biasing using a source meter (Keithley 2612B). Spectral imaging was carried out using an infrared objective with a 0.3 NA. An uncooled microbolometer focal plane array (480 × 480 pixels) was used for detection. All measurements were background-referenced to an Au mirror.

## Funding and Acknowledgements

This work was supported by DARPA via subcontract from Physical Sciences Inc. (Federal Award HR0011-22-C-0068). We also acknowledge support from the UW-Madison Grainger Institute of Engineering, and the Wisconsin Center for Semiconductor Thermal Photonics, with funding from the Wisconsin Alumni Research Foundation (WARF) via the Research Forward Initiative. We also acknowledge the use of facilities and instrumentation in the Wisconsin Center for Nanoscale Technology, supported by the Wisconsin Materials Research Science and Engineering Center (NSF DMR-2309000). T. Kumar acknowledges support from the NASA FINESST grant (80NSSC25K0314).

## Conflict of Interest

The authors declare no conflicts of interest.

## Data Availability Statement

The data that support the findings of this study are available from the corresponding author upon reasonable request.

*Supplementary Information*

# Bandwidth-tunable thermal emitter via dynamic resonator coupling and loss

Jin-Woo Cho[1,*], Tanuj Kumar[1], Joseph A. Andrade[2], Shenwei Yin[1], Samir Rosas[3], Filiz Yesilköy[3], Dung T. Quach[2], Albert G. Wright[2], Colin M. Hessel[2], David Woolf[2], and Mikhail A. Kats[1]

[1]*Department of Electrical and Computer Engineering, University of Wisconsin-Madison, Madison, Wisconsin 53706, USA*
[2]*Physical Sciences Inc., Andover, Massachusetts 01810, USA*
[3]*Department of Biomedical Engineering, University of Wisconsin-Madison, Madison, Wisconsin 53706, USA*

**Corresponding author: cho252@wisc.edu*

This document includes:

Supplementary Sections I–V

Supplementary Figures S1–S10

Supplementary Movie S1

**Supplementary Section I. Calculated reflectivity, transmittivity, and emissivity of free-standing $VO_2$ films**

We calculated the reflectivity, transmittivity, and emissivity (equal to 1 − reflectivity − transmittivity by Kirchhoff's law) of free-standing $VO_2$ films as a function of the $VO_2$ thickness ($t_V$), in both states, using the transfer-matrix method (TMM).

In the insulating phase (**Figs. S1a–S1d**), the $VO_2$ film is mostly transparent below ~12 μm, and thus the optical response is mostly governed by reflection and transmission regardless of $t_V$, while the emissivity remains very low. Note that for the largest thicknesses and shortest wavelengths, we also see the onset of Fabry-Pérot (FP) effects.

When the $VO_2$ film is in its metallic phase (**Figs. S1e–S1h**), its optical response is divided into two regimes relative to the penetration depth ($\delta \sim 130$ nm at $\lambda = 8$ μm). When the film thickness approaches or exceeds the penetration depth, the incident field decays within the highly lossy metallic $VO_2$ film. In contrast to the insulating phase, no FP interference fringes appear, the transmission is suppressed, and strong front-surface reflection governs the response. When the film is far thinner than the penetration depth (~$\delta/5$), the front-surface reflection is strongly reduced, and the field can penetrate the entire layer, enabling transmission (**Fig. S1g**). The field gets partially absorbed, enhancing the emissivity in this thickness regime. However, the emissivity of a thin free-standing film cannot exceed 0.5, and enhancing it requires a back reflector and a cavity that return the wave to the lossy layer for multiple passes, as introduced in **Fig. 1a** of the main text.

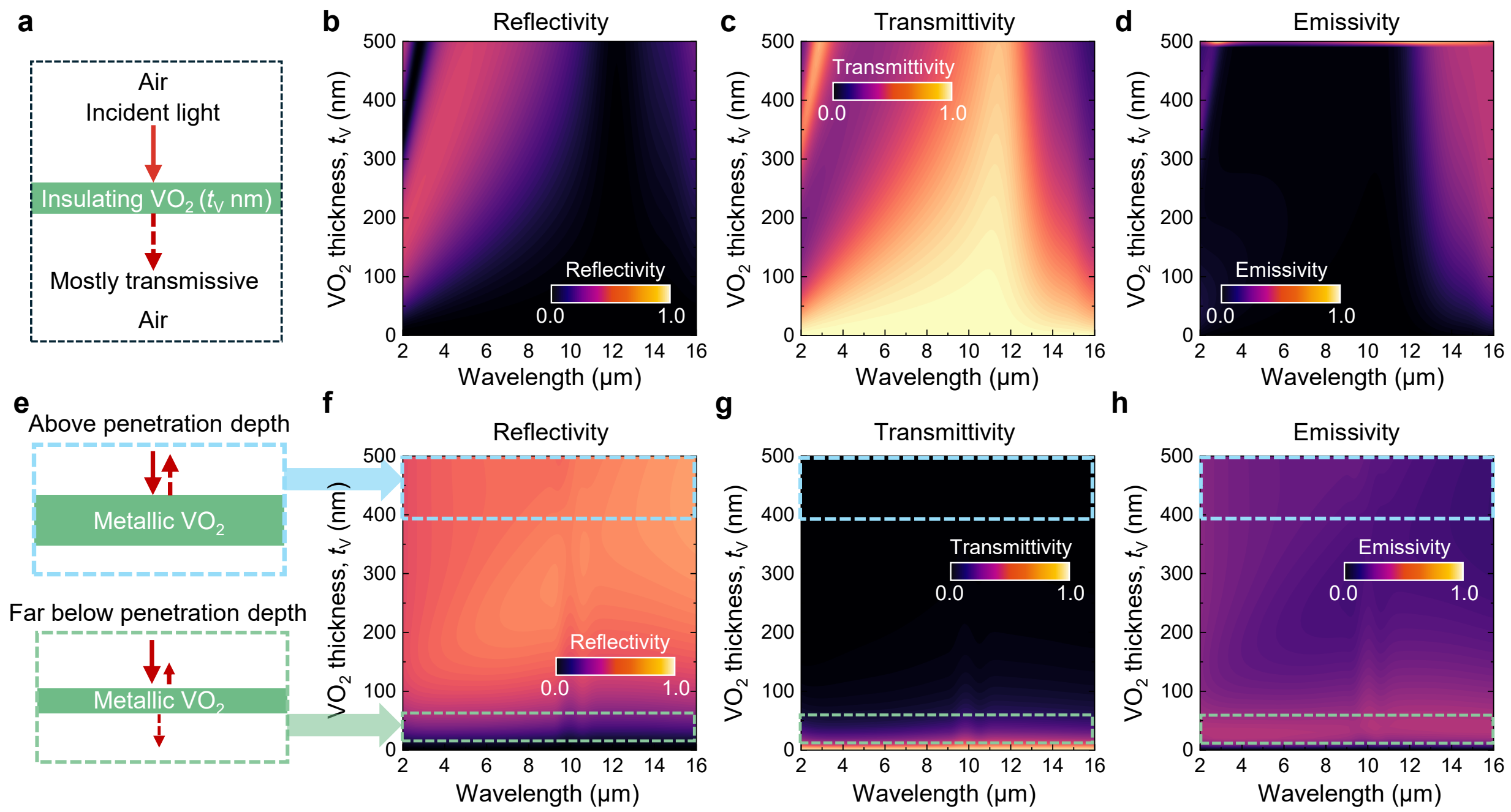

**Supplementary Figure S1. Calculated optical responses of free-standing $VO_2$ films in the insulating and the metallic phases.** (**a**) Schematic of a free-standing $VO_2$ film of thickness, $t_V$, in the insulating phase. (**b–d**) Transfer-matrix method (TMM) calculation results of (b) reflectivity, (c) transmittivity, and (d) emissivity for the $VO_2$ film in its insulating phase as a function of $t_V$. (**e**) Schematics of free-standing metallic $VO_2$ films when (upper) $t_V$ is far above the penetration depth, $\delta$, and (lower) far below it. The two regimes are marked as dashed boxes in the corresponding regions. (**f–h**) TMM calculation results of (f) reflectivity, (g) transmittivity, and (h) emissivity for the $VO_2$ film in its metallic phase as a function of $t_V$.

For a quantitative comparison of the optical responses as a function of $t_V$ in the metallic phase, we plotted the spectra at representative values of $t_V$ in **Fig. S2**. With decreasing $t_V$, the reflectivity decreases significantly, from ~0.7 at $t_V$ = 400 nm to ~0.15 at $t_V$ = 20 nm, enhancing the emissivity from ~ 0.3 at $t_V$ = 400 nm to ~0.5 at $t_V$ = 20 nm.

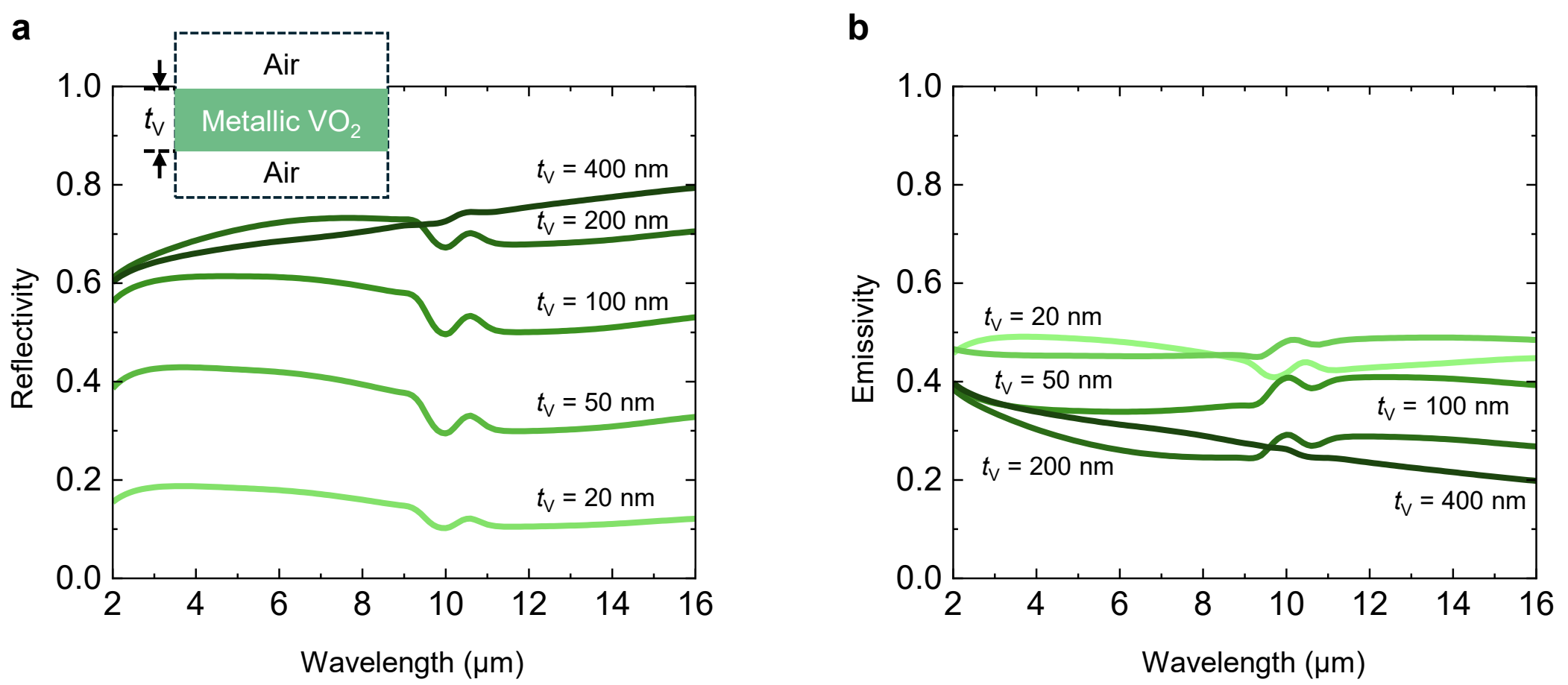

**Supplementary Figure S2. Calculated reflectivity and emissivity spectra for free-standing $VO_2$ films in the metallic phase.** (**a**) Reflectivity and (**b**) emissivity spectra at the representative values of $t_V$ selected from Figs. S1f and S1h.

## Supplementary Section II. Transmission-line analysis

A planar multilayer at normal incidence can be analyzed using a transmission-line (TL) model. A single input impedance $Z_{in}$ can be calculated for the entire multilayer stack, and can then be compared with the free-space impedance, $Z_0$ $(= \sqrt{(\mu_0/\epsilon_0)} \approx 377\ \Omega)$. The metallic $VO_2$ film is treated as a shunt sheet element with a complex sheet impedance $Z_{VO2} = 1/(\tilde{\sigma} t_V)$, where $\tilde{\sigma} = -i\omega\epsilon_0(\tilde{\epsilon} - 1)$ is the complex optical conductivity obtained from the measured permittivity. At the 20 nm thickness used here, the propagation phase across the film is small $|\beta_V t_V| \approx 0.1$, where $\beta_V = 2\pi\tilde{n}_V/\lambda$, and the film remains thinner than its penetration depth across 6–14 μm. The accuracy of the resulting shunt-sheet approximation is verified numerically in **Fig. S3**.

We verified the shunt-sheet treatment by simplifying the model to a free-standing metallic $VO_2$ film and comparing the optical responses calculated via TMM and the TL model as a function of $t_V$ (**Fig. S3a**). In the TL model, the $VO_2$ film is represented as a shunt sheet loaded by free space on the transmission side. The corresponding input impedance $Z_{in}$ is

$$Z_{in} = \frac{Z_{VO2} Z_0}{Z_{VO2} + Z_0}, \qquad \textbf{Eq. S1}$$

The reflection and transmission coefficients are $\Gamma = (Z_{in} - Z_0)/(Z_{in} + Z_0)$ and $\tau = 1 + \Gamma$, respectively. Thus, the absorptivity is $A = 1 - |\Gamma|^2 - |\tau|^2$. The absorptivity difference between the two methods $|A_{TL} - A_{TMM}|$ is below 1% for $t_V \leq 80$ nm across 6–14 μm (**Fig. S3b**). Therefore, the shunt-sheet approximation is quantitatively justified up to 80 nm, and remains accurate to within 2% even up to 100 nm.

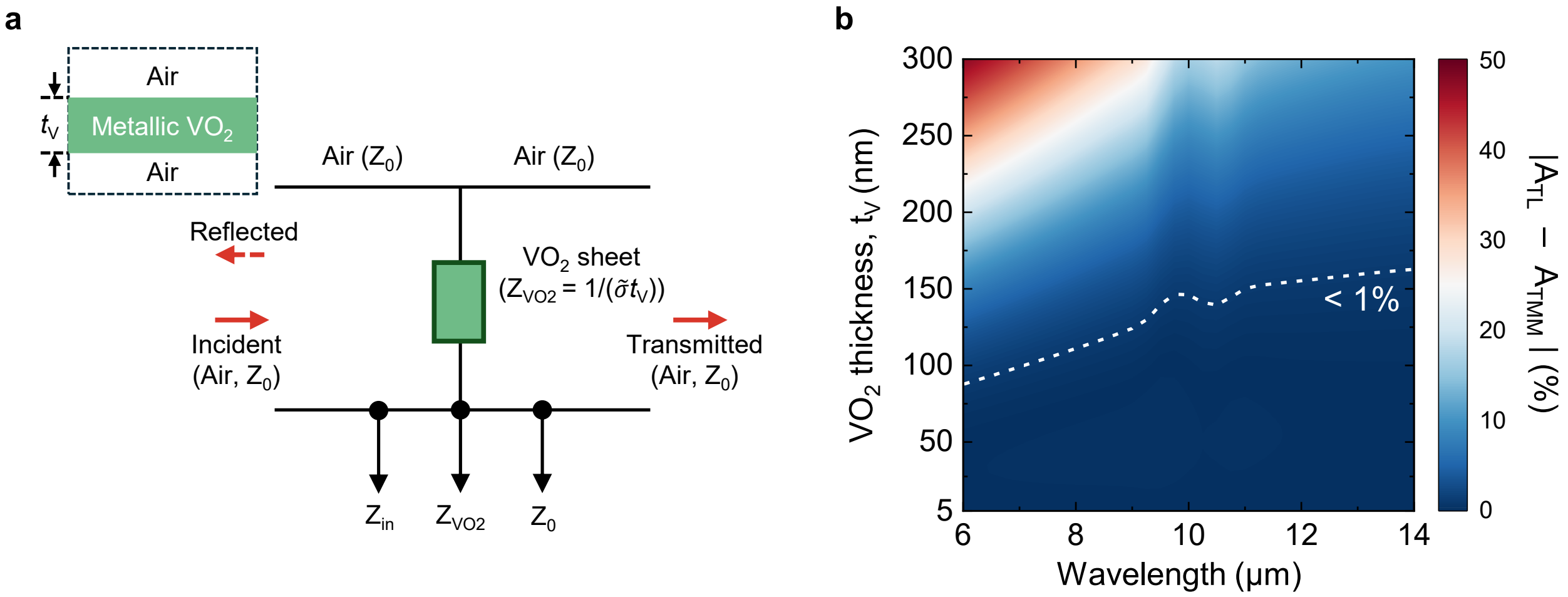

**Supplementary Figure S3. Validation of the transmission-line shunt-sheet approximation for a free-standing metallic $VO_2$ film.** (**a**) Schematic shows the free-standing metallic $VO_2$ film and its transmission-line (TL) model, in which the $VO_2$ film is represented as a shunt-sheet element. (**b**) Absolute absorptivity difference between the TL model and transfer-matrix method (TMM), $|A_{TL} - A_{TMM}|$, as a function of wavelength and $t_V$. Dashed line indicates the 1% deviation contour.

The full stack is solved by successive impedance transformation starting from the gold (Au) mirror (**Fig. S4a**). The $SiO_2$ layer and the Si spacer are modeled as cascaded transmission-line sections, each with characteristic impedance $Z_m = Z_0/\tilde{n}_m$ and propagation phase $\beta_m t_m = 2\pi\tilde{n}_m t_m/\lambda$, where $\tilde{n}_m$ and $t_m$ are the refractive index and thickness of each layer ($m$ = $SiO_2$ and spacer). The optically thick Au mirror is approximated as a short circuit $Z_{sub} = 0$, which eliminates transmission. Here $Z_1$, $Z_2$, and $Z_{in}$ denote the transformed impedance at each plane looking toward the mirror, whereas $Z_{SiO2}$ and $Z_{spacer}$ are the component impedances of the corresponding layers, and $Z_{VO2}$ is the $VO_2$ sheet impedance. Starting from $Z_{sub} = 0$, the spacer transforms the short into $Z_1 = -iZ_{spacer}\tan(\beta_{spacer}t_{spacer})$ at the $VO_2$ plane. The $VO_2$ sheet is then added in parallel through $1/Z_2 = 1/Z_1 + 1/Z_{VO2}$, and the $SiO_2$ layer transforms $Z_2$ into the input impedance $Z_{in}$ at the air interface as follows:

$$Z_{in} = Z_{SiO2}\frac{Z_2 - iZ_{SiO2}\tan(\beta_{SiO2}t_{SiO2})}{Z_{SiO2} - iZ_2 tan(\beta_{SiO2}t_{SiO2})} \quad \textbf{Eq. S2}$$

The reflection coefficient follows from $\Gamma = (Z_{in} - Z_0)/(Z_{in} + Z_0)$. Because there is no transmission, the absorptivity ($A$) is $A = 1 - |\Gamma|^2$, which equals the emissivity by Kirchhoff's law. Finally, the resulting impedance mismatch between the full stack $Z_{in}$ and the free space $Z_0$, $|Z_{in} - Z_0|$, is plotted as a function of wavelength and $t_V$ in **Fig. S4b**. At each wavelength we identify the thickness minimizing $|Z_{in} - Z_0|$, Min$|Z_{in} - Z_0|$, and this contour is plotted together in **Fig. S4b** and the TMM-calculated emissivity map shown in **Fig. 2e** of the main text. The analysis exhibits $t_V \sim 20$ nm as the thickness maintaining the near impedance-matching condition over the widest spectral range, mostly 7–11 µm. Consistently, the average emissivity over λ = 8–14 µm, weighted by the 300-K and 340-K blackbody spectra, provides its maximum near $t_V \sim 20$ nm (**Fig. S5**).

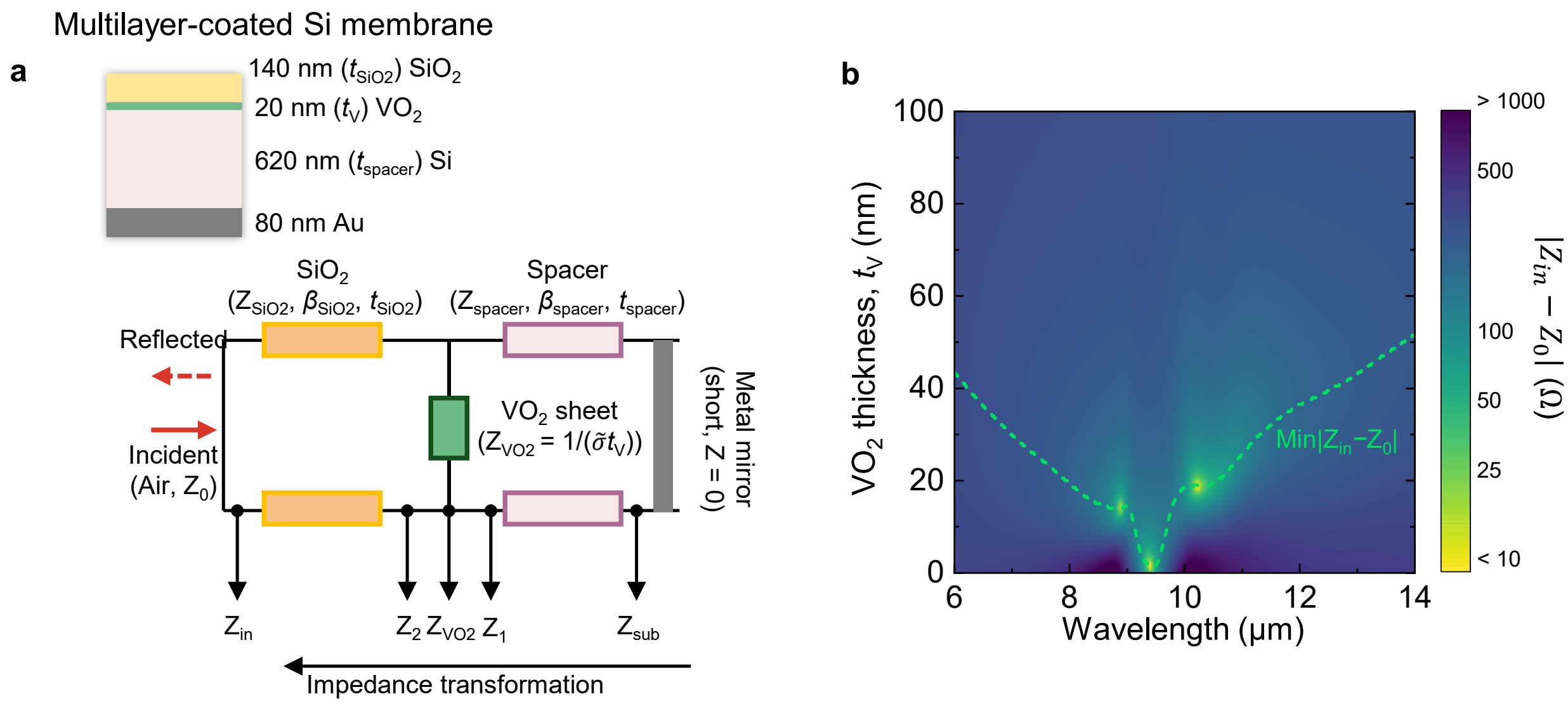

**Supplementary Figure S4. Impedance mismatch analysis of the full-stack transmission-line model.** (**a**) Schematic shows the full-stack structure and its corresponding transmission-line (TL) model, in which the $VO_2$ film is represented as a shunt-sheet element. (**b**) Impedance mismatch between the input impedance of the full stack $Z_{in}$ and the free space $Z_0$, $|Z_{in} - Z_0|$, as a function of wavelength and $t_V$. Dashed line indicates the thickness minimizing $|Z_{in} - Z_0|$ at each wavelength, Min$|Z_{in} - Z_0|$, which is also plotted in Fig. 2e.

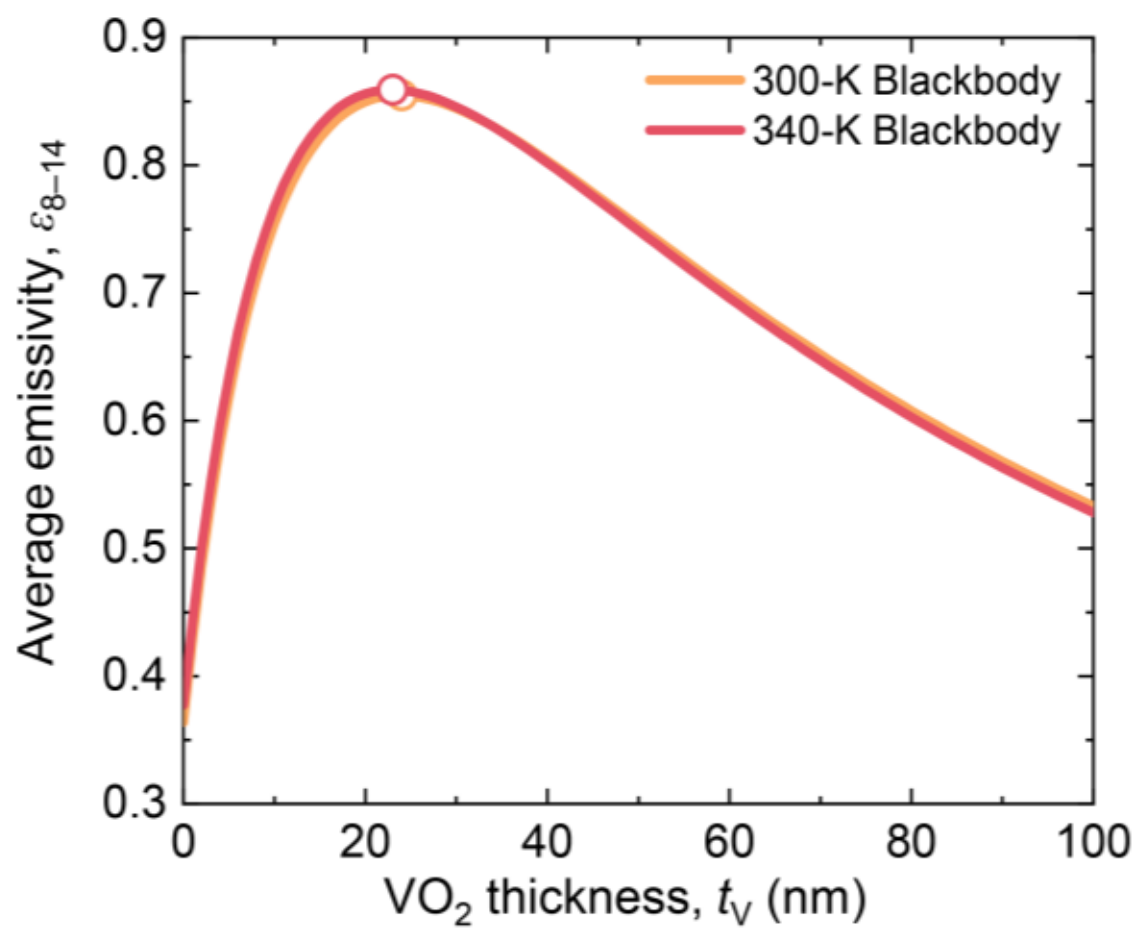

**Supplementary Figure S5. Average emissivity as a function of $VO_2$ thickness.** The average emissivity is calculated for the full stack in Fig. S4 when the $VO_2$ film is in its metallic phase and weighted by the 300-K and 340-K blackbody spectra over λ = 8–14 µm. Symbols mark the thickness of maximum average emissivity for each temperature.

## Supplementary Section III. Ge compensation layer for the spacer thickness deficit

The design in **Fig. 2** requires a Si spacer thickness of ~620 nm, whereas standard commercially available c-Si membranes we were able to obtain are 500 nm thick. To compensate for this "thickness deficit", we added a Ge layer on the backside. Because Ge is an infrared transparent, high-index ($n$ ~ 4) material, the additional layer can be kept thin (**Fig. S6a**, upper). The calculated emissivity of the full stack in the insulating state as a function of Ge thickness (keeping c-Si fixed at 500 nm) shows that the narrowband peak at λ ~ 9.4 µm is achieved when the Ge thickness is > 40 nm (**Fig. S6a**, lower). The angle-resolved emissivity plots confirm that the 500-nm Si with the 50-nm-thick Ge layer shows nearly the same results as the original 620-nm Si spacer design in both phases.

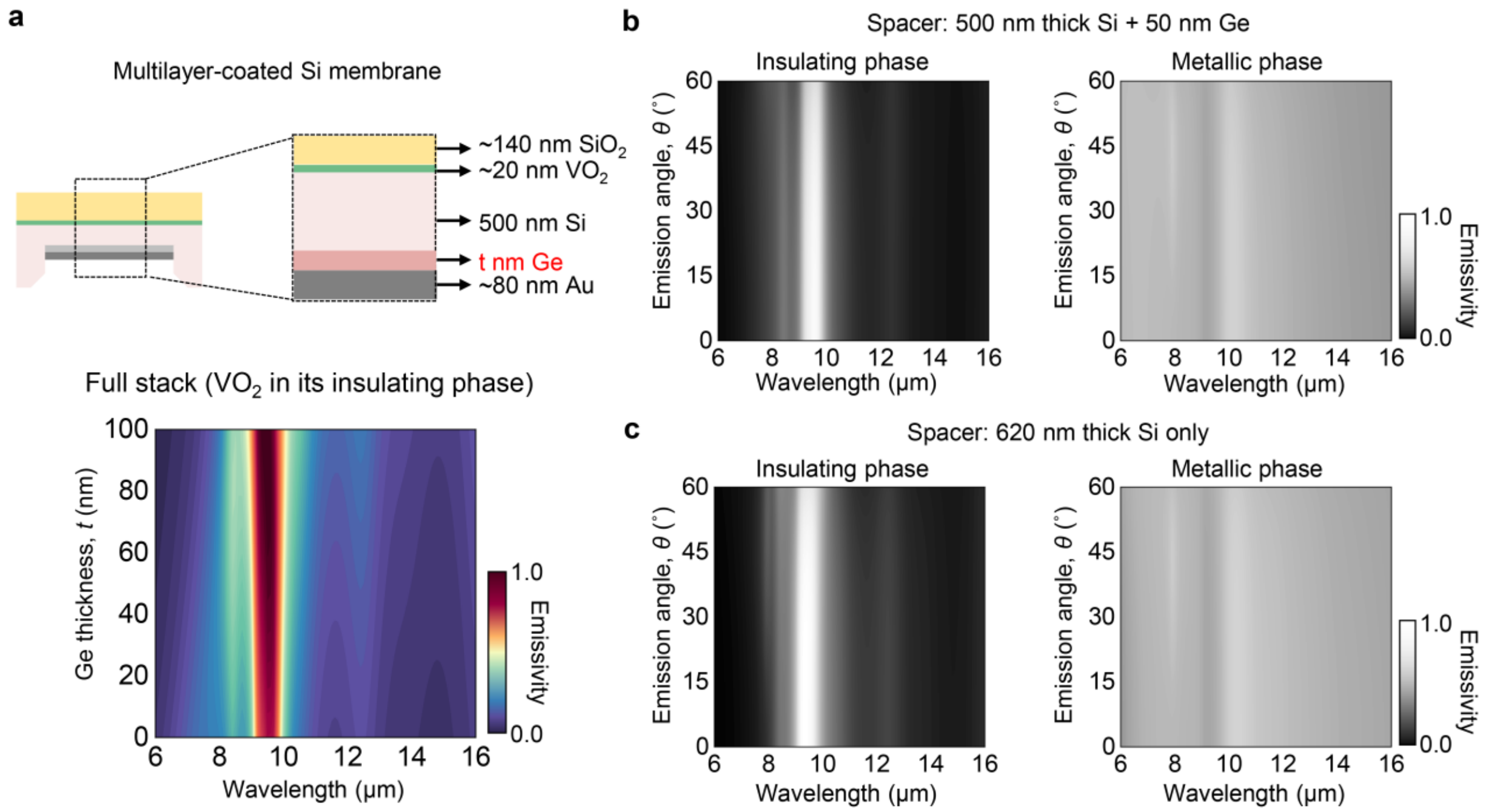

**Supplementary Figure S6. Ge compensation layer design.** (**a**) (Upper) The multilayer coated Si membrane with Ge layer on the bottom side. (Lower) Calculated emissivity of the full stack in the insulating phase as a function of Ge thickness, for Si fixed at 500 nm. (**b**, **c**) Angle-resolved emissivity plots of (b) the fabricated design with a 500 nm Si spacer and a 50 nm Ge layer and (c) the original design with a 620 nm Si spacer, in the insulating and metallic phases.

## Supplementary Section IV. Electrically switchable bandwidth-tunable thermal-emitting device

For electrical switching, we deposited a gold microheater onto the top surface of the membrane chip by e-beam evaporation through a metal shadow mask (**Fig. S7**). We then integrated the chip into a printed circuit board (PCB) and connected the heater electrodes to the PCB pads via aluminum wire bonding.

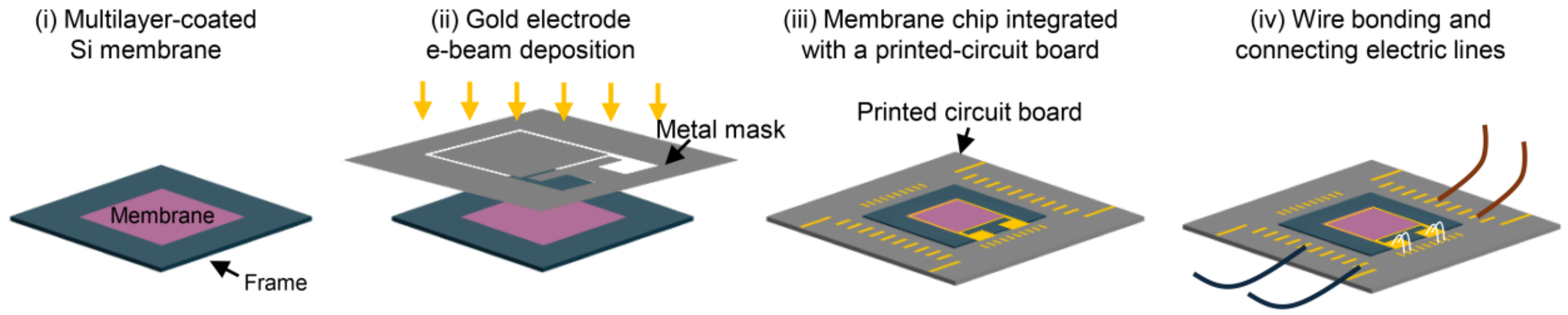

**Supplementary Figure S7. Fabrication of the electrically switchable device.** Schematics show the process of integrating the heater onto our multilayer-coated membrane chip and then onto a PCB, and connecting the electric lines.

Then we compared the emissivity spectra under electrical biasing with those measured using an external heating stage before the PCB integration (**Fig. S8**). The external heating results from 30 to 90 °C (**Fig. S8a**) and the electrical biasing results from 0 to 26 mA (**Fig. S8b**) show nearly the same emissivity for both the narrowband and broadband states, as the $VO_2$ layer transitions from its insulating to metallic phase.

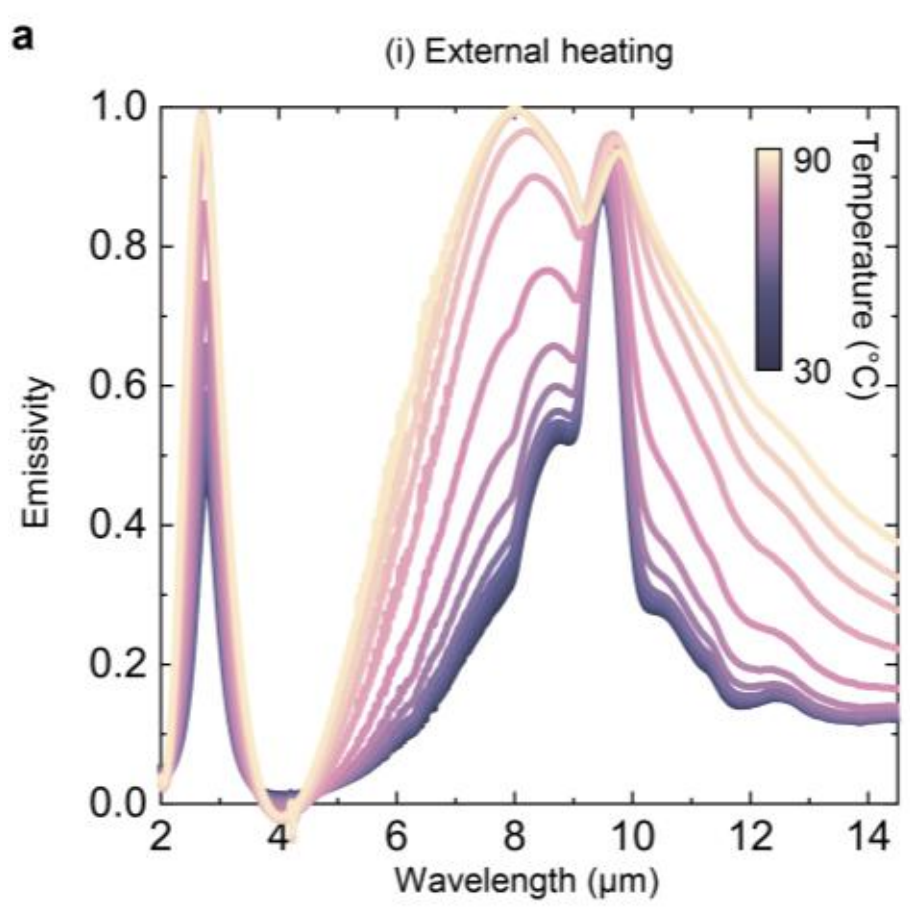

**Supplementary Figure S8. Emissivity spectra measured under the two heating configurations.** Emissivity spectra measured (**a**) using the external heating stage as a function of temperature and (**b**) under electrical biasing as a function of current.

## Supplementary Section V. Hyperspectral imaging and extension to other polar dielectrics

To motivate the concept of infrared tags and labels from **Fig. 5** in the main text, we performed hyperspectral reflectance imaging of the fabricated device in **Fig. 3a** over a 2 mm × 2 mm field of view (see **Methods** in the main text for details). We obtained the emissivity as 1 − reflectivity by Kirchhoff's law. We tuned the $VO_2$ to its insulating and metallic phases by applying electrical currents of 0 and 26 mA. The emissivity spectra in **Fig. S9a** reproduce the spectral features of the FTIR measurements in **Figs. 4b** and **S8**. The emissivity at the resonant wavelength ($\lambda$ = 9.4 μm) remains nearly unchanged across the IMT, whereas that at the non-resonant wavelength ($\lambda$ = 8.0 μm) changes markedly (**Fig. S9b**). In the metallic phase, the two wavelengths show similarly high emissivity. This is particularly striking when looking at the hyperspectral images of the suspended structures, in **Fig. S9b**.

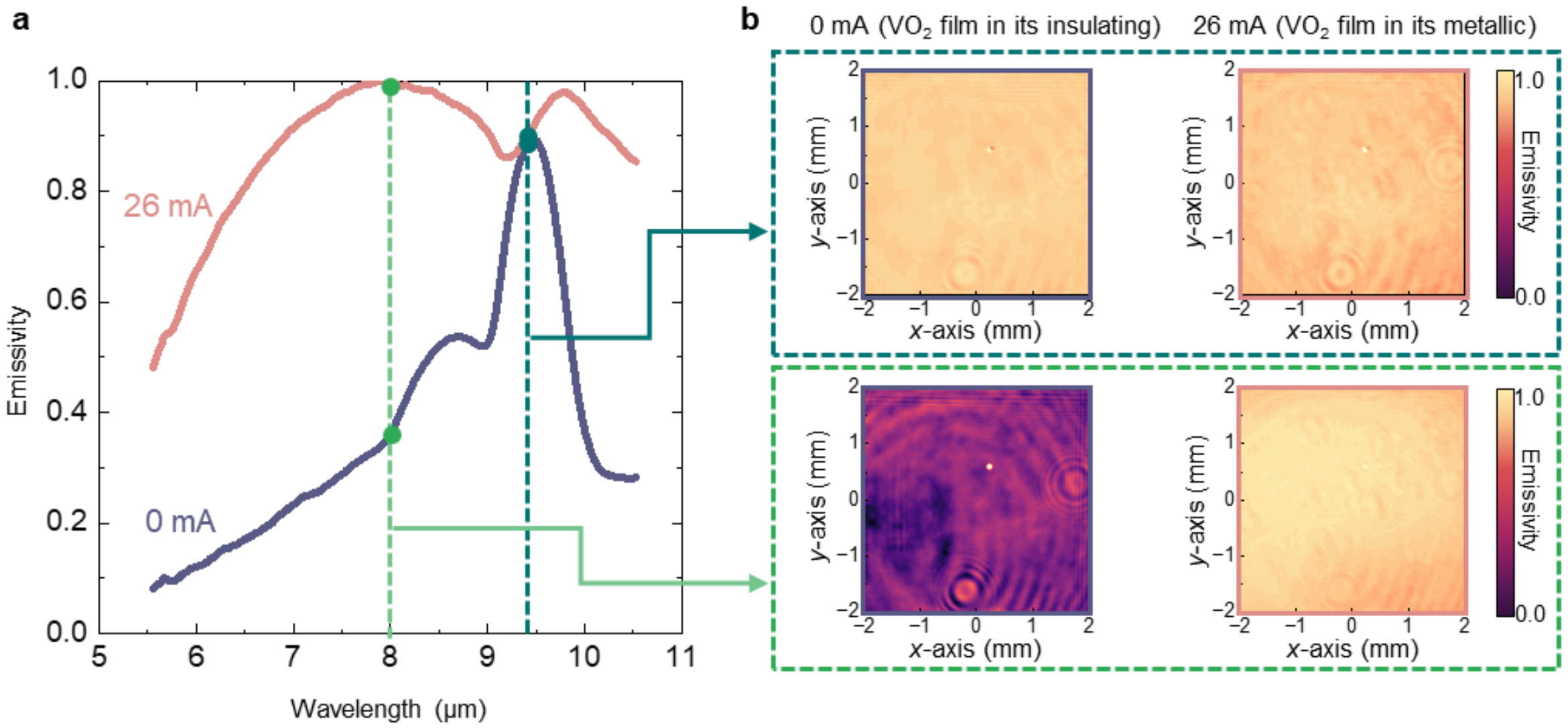

**Supplementary Figure S9. Hyperspectral reflection imaging.** (**a**) Emissivity spectra at 0 mA (insulating phase) and 26 mA (metallic phase). The dashed lines mark the non-resonant (8.0 μm) and resonant (9.4 μm) wavelengths. (**b**) Emissivity maps at 9.4 μm (upper) and 8.0 μm (lower) for 0 mA (left) and 26 mA (right) over the 4 × 4 $mm^2$ field of view.

To change the narrowband emission wavelength, we can replace the $SiO_2$ layer with other polar dielectrics, for example $Si_3N_4$ or $Al_2O_3$, while keeping the 20-nm-thick $VO_2$ layer as before (**Fig. S10**). For each polar dielectric, we re-optimized the thicknesses of both the polar-dielectric and spacer, to align the FP mode of the cavity with the vibrational resonance of the dielectric, as shown in **Figs. S10a** and **S10d**.

The calculated angle-resolved emissivity plots confirm the bandwidth tuning capability for both stacks comprising the different polar dielectrics. When the $VO_2$ is in its insulating phase, the structures with $Si_3N_4$ and $Al_2O_3$ layers show narrowband emissivity at their vibrational resonances at $\lambda \sim 12$ µm and $\lambda \sim 17$ µm, respectively (**Figs. S10b** and **S10e**). When the $VO_2$ transitions to its metallic phase, both structures show broadband emissivity (**Figs. S10c** and **S10f**), though the exact range of the broadband emissivity can also vary. Regardless of the $VO_2$ phase, the emissivity spectrum of each structure remains nearly angle-insensitive up to 60° away from the normal.

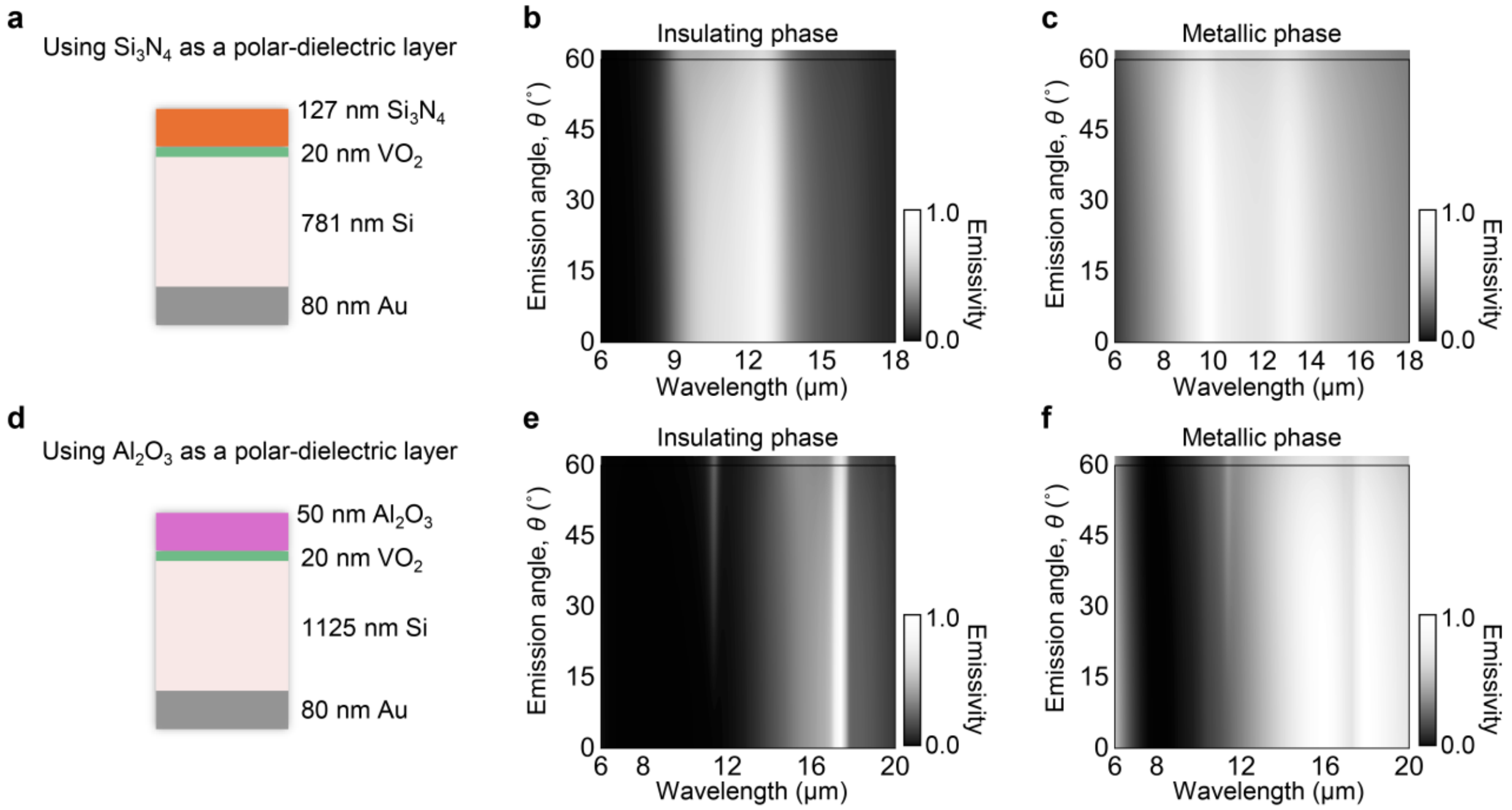

**Supplementary Figure S10. Design with other polar dielectrics.** (**a**, **d**) Structure using (a) $Si_3N_4$ and (d) $Al_2O_3$ as the polar-dielectric layer. Calculated angle-resolved emissivity plots when the $VO_2$ is in its (**b**, **e**) insulating and (**c**, **f**) metallic phases.